\documentclass[a4paper,11pt]{article}
\pdfoutput=1 

\usepackage{jheppub}
\usepackage[T1]{fontenc} 

\usepackage{tikz}
\usetikzlibrary{patterns}
\usetikzlibrary{snakes}

\newcommand{\gen}[1]{\mathbf{#1}}
\newcommand{\ket}[1]{|#1\rangle}
\newcommand{\sech}{\text{sech}}
\newcommand{\arcsinh}{\text{arcsinh}}
\newcommand{\arccosh}{\text{arccosh}}

\newcommand{\bal}{\begin{equation}\begin{aligned}}
\newcommand{\eal}{\end{aligned}\end{equation}}

\title{\boldmath Massless S matrices for AdS3/CFT2}

\author[a,1]{Sergey Frolov,%
\note{Correspondent fellow at the Steklov Mathematical Institute, Moscow.}}
\author[b,c,2]{Alessandro Sfondrini%
\note{IBM Einstein Fellow.}}

\affiliation[a]{School of Mathematics and Hamilton Mathematics Institute,\\
Trinity College, Dublin 2, Ireland}
\affiliation[b]{Institute for Advanced Study,\\
Einstein Drive, Princeton, New Jersey, 08540 USA}
\affiliation[c]{Dipartimento di Fisica e Astronomia, Universit\`a degli Studi di Padova,\\
\& Istituto Nazionale di Fisica Nucleare, Sezione di Padova,\\
via Marzolo 8, 35131 Padova, Italy}

\emailAdd{frolovs@maths.tcd.ie}
\emailAdd{alessandro@ias.edu}

\abstract{The AdS3/CFT2 correspondence features massless non-relativistic modes on the string worldsheet in lightcone gauge. We study in detail these excitations and highlight how they naturally split between chiral (left-moving) and anti-chiral (right-moving) representations. In light of this split we discuss the constraints on the two-particle worldsheet S matrix imposed by braiding and physical unitarity, parity, time-reversal, and crossing invariance. We also comment on the implication of this split in the mirror kinematics.}

\begin{document} 
\maketitle
\flushbottom

\section{Introduction and main results}
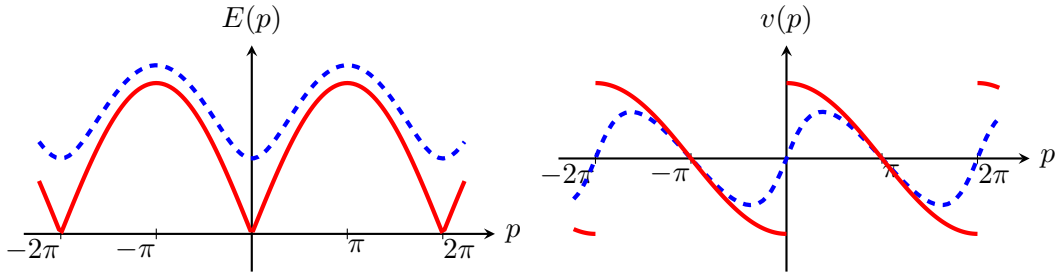
\begin{figure}
	\centering
\begin{tikzpicture}[>=stealth,xscale=0.4]
    \def\pigr{3.14159}  
    \draw[thick, ->] (-7.5,0) -- (8,0) node[right] {$p$};
    \draw[thick, ->] (0,-0.5) -- (0,2.5) node[above] {$E(p)$};
\foreach \pos/\et/\anch in {-2*\pigr/-2\pi/north east,-\pigr/-\pi/north east,\pigr/\pi/north west,2*\pigr/2\pi/north west}
    \draw[shift={(\pos,0)}] (0pt,2pt) -- (0pt,-2pt) node[anchor=\anch,inner sep=0pt] {$\et$};
    \draw[ultra thick,blue, dashed,domain=-7:7,samples=100]
    plot (\x,{sqrt(1+4*sin(\x r*0.5)*sin(\x r*0.5))});
    \draw[ultra thick,red,domain=-7:7,samples=150]
    plot (\x,{sqrt(4*sin(\x r*0.5)*sin(\x r*0.5))});
    \end{tikzpicture}
    \begin{tikzpicture}[>=stealth,xscale=0.4]
    \def\pigr{3.14159}  
    \draw[thick, ->] (-7.5,0) -- (8,0) node[right] {$p$};
    \draw[thick, ->] (0,-1.5) -- (0,1.5) node[above] {$v(p)$};
\foreach \pos/\et/\anch in {-2*\pigr/-2\pi/north east,-\pigr/-\pi/north east,\pigr/\pi/north west,2*\pigr/2\pi/north west}
    \draw[shift={(\pos,0)}] (0pt,2pt) -- (0pt,-2pt) node[anchor=\anch,inner sep=0pt] {$\et$};
    \draw[ultra thick,blue, dashed,domain=-7:7,samples=100]
    plot (\x,{sin(\x r)/(sqrt(3-2*cos(\x r)))});
    \draw[ultra thick,red,domain=0:6.2832,samples=80]
    plot (\x,{cos(\x r*0.5))});
    \draw[ultra thick,red,domain=-6.2832:0,samples=80]
    plot (\x,{-cos(\x r*0.5))});    \draw[ultra thick,red,domain=-7:-6.2832,samples=80]
    plot (\x,{cos(\x r*0.5))});
    \draw[ultra thick,red,domain=6.2832:7,samples=80]
    plot (\x,{-cos(\x r*0.5))});
    \end{tikzpicture}
	\caption{
	kinematics of massive and massless modes in $AdS_3\times S^3\times T^4$. On the left, we plot the  dispersion relation $E(p)$ for massive particles (blue dashed line) and for massless ones (red solid line); the latter dispersion has cusps at $p=0\,$mod$\,2\pi$. On the right, we do the same for the group velocity $v(p)$, which form for massless particles is discontinuous  $p=0\,$mod\,$2\pi$.
	}
\label{fig:dispersion}
\end{figure}

The AdS3/CFT2 correspondence is a particularly rich holographic setup~\cite{Maldacena:1997re,Witten:1998qj,Gubser:1998bc}. Its most supersymmetric instances features backgrounds of the form $AdS_3\times S^3\times M$ where $M$ could be the four-torus $T^4$, a $K3$ surface, or the product of a sphere and a circle, $S^3\times S^1$. Among these, the case of $T^4$ is perhaps the simplest and it is the one on which we shall mostly focus. On the one hand, it is possible to support this background by Neveu-Schwarz-Neveu-Schwarz (NSNS) fluxes only, in which case it is possible to understand it in terms of a Wess-Zumino-Witten (WZW) model~\cite{Maldacena:2000hw}, see also refs.~\cite{Giribet:2018ada,Eberhardt:2018ouy,Eberhardt:2021vsx} for recent work in this direction. On the other hand, it is also possible to consider backgrounds supported by Ramond-Ramond (RR) fluxes, or by a combination of RR and NSNS fluxes. These cases do not seem to be amenable to worldsheet-CFT approaches, nor can they be approached by considering their dual CFT, which is thus far unknown. However, they can be studied by using integrability techniques on the string worldsheet in lightcone gauge, see~\cite{Sfondrini:2014via} for a review. The case of pure-RR background, which is the one farthest from the WZW model, is the simplest in the context of worldsheet integrability.%
\footnote{The case of pure-NSNS backgrounds can also be thoroughly understood by integrability~\cite{Baggio:2018gct,Dei:2018mfl} (including when the WZW level has the special value $k=1$ \cite{Eberhardt:2018ouy,Sfondrini:2020ovj}), while the mixed-flux theories show signs of being integrable but appear to have a much more complicated dynamics~\cite{Hoare:2013pma,Hoare:2013lja, Lloyd:2014bsa,Babichenko:2014yaa}.}
In particular, its dynamics closely resembles that of strings on $AdS_5\times S^5$, which is by now very well-understood in terms of integrability, at least for what concerns the spectral problem,%
\footnote{The computation of three-~\cite{Basso:2015zoa} and higher-point~\cite{Eden:2016xvg,Fleury:2016ykk} correlation functions by integrability saw significant  advances recently, including in AdS3/CFT2~\cite{Eden:2021xhe}, but it is far from being completely understood.}
see~\cite{Arutyunov:2009ga,Beisert:2010jr} for reviews.
Still a major difference between the case of AdS3/CFT2 and AdS5/CFT4 is the dispersion relation for worldsheet excitations in lightcone gauge,
\begin{equation}
    E(p,M) = \sqrt{M^2+4h^2\sin^2\Big(\frac{p}{2}\Big)}\,.\,
\end{equation}
where $h>0$ is the coupling constant.
We will review later how this is quite a robust feature of the integrable model, as it follows from representation theory of the lightcone symmetry algebra~\cite{Borsato:2012ud,Borsato:2013qpa}. Crucially, and unlike in AdS5/CFT4, in AdS3/CFT2 we may have $M=0$~\cite{Borsato:2014exa}. It immediately follows that the dispersion relation is not analytic around $p=0$, like for relativistic massless modes in $1+1$ dimensions. However, unlike the relativistic case, the group velocity of the massless modes
\begin{equation}
    v(p) =\frac{\partial}{\partial p} E(p,0) = \pm h\,\cos\Big(\frac{p}{2}\Big)\,,
\end{equation}
is not constant, see also figure~\ref{fig:dispersion}. The main aim of this work is to study the kinematics of these non-relativistic massless excitations (and of their integrable S~matrix~\cite{Borsato:2014exa,Borsato:2014hja}), being especially careful about the non-analytic behaviour around $p=0$.

An important insight to study the scattering of massless modes was obtained by Fontanella and Torrielli~\cite{Fontanella:2019baq}. They proposed a rapidity variable~$\gamma(p)$ for massless particles, similar to the one discussed in AdS5/CFT4 by Beisert, Hernandez and Lopez~\cite{Beisert:2006ib}, in terms of which the massless-massless scattering greatly simplifies. Indeed, the part of the scattering matrix which is fixed by symmetries is of difference form, when expressed in terms of such a rapidity. Unfortunately however, the change of variable~$\gamma(p)$ needed to introduce such a rapidity is not analytic in terms of the momentum. This begs the question of whether the scattering matrix is an analytic function of $\gamma$ for any values of the momentum, or whether we must distinguish different regions for~$p$ (and which ones). This somewhat technical but important point has not been explored so far in the literature.%
\footnote{%
Some aspects of this issue were discussed in appendix~C of~\cite{Majumder:2021zkr} in the context of the study of zero-momentum excitations of the theory; we thank the anonymous referee for pointing this out to us.
}

We shall see that, when we express the S~matrix in terms of~$\gamma$, we need to distinguish the case where $v(p)$ is positive (we will call these ``chiral'' particles) from the one where it is negative (``anti-chiral''). We then check that our construction is compatible with the expected symmetries of the worldsheet theory, such as physical unitarity, parity, time-reversal and crossing symmetry, as well as with features which we expect in a theory with factorised scattering such as braiding unitarity and the Yang-Baxter equation. We also derive the constraint that such symmetries put on the ``dressing factors'' of the theory --- the parts of the scattering matrix which cannot be entirely fixed by symmetry but are still partially constrained by the aforementioned symmetries. While for $AdS_3\times S^3\times T^4$ there exists a proposal for the dressing factors~\cite{Borsato:2013hoa,Borsato:2016xns}, the analysis that we carry out in this paper will be useful to review and amend that proposal, which we will do elsewhere~\cite{Frolov:2021fmj}.

Finally, we will briefly discuss the kinematics of the so-called mirror theory. This is a model related to the original theory on the string worldsheet by a double-Wick rotation, which can also be understood in terms of an analytic continuation in a suitable rapidity variable~\cite{Arutyunov:2007tc}. The study of this theory is crucial in order to formulate the ``mirror'' thermodynamic Bethe ansatz (TBA)~\cite{Zamolodchikov:1989cf} of a non-relativistic model such as the one at hand, which in turn is necessary to describe finite-volume effects in the theory's spectrum~\cite{Ambjorn:2005wa}. Formulating the mirror TBA for AdS3/CFT2 in presence of Ramond-Ramond flux is an outstanding open problem to which we plan to return soon~\cite{Frolov:2021bwp} (the case of pure-NSNS flux was  much simpler, and it was discussed in~\cite{Dei:2018mfl,Dei:2018jyj}).%
\footnote{%
Some attempts at formulating the mirror TBA for RR backgrounds have been undertaken in the literature, but they did not account for all excitations of the model and their correct kinematics.
}
In this paper we will limit ourselves to studying the mirror kinematics for massless particles (that of massive particles follows straightforwardly from~\cite{Arutyunov:2007tc}) and understand whether particles of different chirality lead to distinct scattering matrices. We find that, like in the string-worldsheet theory, this is the case, due to the non-analytic behaviour. In the mirror case, rather than appearing around $E=p=0$, this appears when the mirror energy and momentum vanish, $E_m=p_m=0$.

This paper is structured as follows. We start by briefly recalling the representations of the light-cone symmetry algebra that fix the AdS3/CFT2 kinematics in section~\ref{sec:symmetries}, with particular attention to the massless representations.%
\footnote{The complete discussion of the representations and the resulting S~matrix is somewhat scattered throughout the literature, unfortunately with varying notations. A recent rather complete summary of the main results can be found in~\cite{Eden:2021xhe}.}
In section~\ref{sec:chiralsplit} we focus on the split between chiral and anti-chiral particles, and we show that they sit in distinct representations. Based on this, in section~\ref{sec:Smatrices} we derive the various S~matrices in the sectors where one or both particles are massless, check their consistency, and derive the constraints on their scalar factors. Finally, in section~\ref{sec:mirror} we discuss the massless kinematics of the mirror model. As the formulae for the S matrices involving one massive and one massless particle are a little bulky, we relegate them to appendix~\ref{app:smatrix}.

\section{The symmetry algebra and its representations}
\label{sec:symmetries}
We start by briefly reviewing the symmetries of the S~matrix which will govern many of its features.
The symmetry algebra of the $AdS_3\times S^3\times T^4$ in the lightcone gauge is given by%
\footnote{Sometimes the central charge are called $\gen{P}$ rather than $\gen{C}$, and $\gen{K}$ rather than $\bar{\gen{C}}$; we do not use that notation to avoid confusion with the momentum, which we will call $\gen{p}$.}
\begin{equation}
\label{eq:bigalgebra}
\begin{aligned}
    \{\gen{Q}^{A},\gen{S}_{B}\} =\delta^A_B\,\gen{H}\,,\qquad
    \{\widetilde{\gen{Q}}_{A},\widetilde{\gen{S}}^{B}\} =\delta_A^B\,\widetilde{\gen{H}}\,,\\
    \{\gen{Q}^{A},\widetilde{\gen{Q}}_{B}\} =\delta^A_B\,\gen{C}\,,\qquad
    \{\gen{S}_{A},\widetilde{\gen{S}}^{B}\} =\delta_A^B\,\bar{\gen{C}}\,.
\end{aligned}
\end{equation}
It is possible to construct the relevant representations of this algebra in terms of a smaller algebra, namely
\begin{equation}
\label{eq:smallalgebra}
    \{\gen{q},\gen{s}\}=\gen{H}\,,\qquad
    \{\tilde{\gen{q}},\tilde{\gen{s}}\}=\widetilde{\gen{H}}\,,\qquad
    \{\gen{q},\tilde{\gen{q}}\}=\gen{C}\,,\qquad
    \{\gen{s},\tilde{\gen{s}}\}=\bar{\gen{C}}\,,\
\end{equation}
by setting
\begin{equation}
    \gen{Q}^1=\gen{q}\otimes\gen{1}\,,\qquad
    \gen{Q}^2=\gen{\Sigma}\otimes\gen{q}\,,\qquad
    \gen{S}_1=\gen{s}\otimes\gen{1}\,,\qquad
    \gen{S}_2=\gen{\Sigma}\otimes\gen{s}\,,
\end{equation}
where $\gen{\Sigma}=(-1)^{\gen{F}}$ is the Fermion sign, and (notice the lowered indices)
\begin{equation}
    \widetilde{\gen{Q}}_1=\tilde{\gen{q}}\otimes\gen{1}\,,\qquad
    \widetilde{\gen{Q}}_2=\gen{\Sigma}\otimes\tilde{\gen{q}}\,,\qquad
    \widetilde{\gen{S}}^1=\tilde{\gen{s}}\otimes\gen{1}\,,\qquad
    \widetilde{\gen{S}}^2=\gen{\Sigma}\otimes\tilde{\gen{s}}\,.
\end{equation}
For simplicity, we will mostly work in terms of the representations of the smaller algebra generated by $\gen{q}, \tilde{\gen{q}}, \gen{s}, \tilde{\gen{s}}$ as this will be sufficient to highlight several important features of the massless representations. However, it is important to bear in mind that the full S~matrix scatters particles which are in representations of the larger algebra; this is relevant when, for instance, we want to appropriately normalise the scattering matrix.
Finally, it is worth noting that for any supercharge~$\gen{q}$ we have
\begin{equation}
\label{eq:sigmaidentity}
    \gen{\Sigma}\,\gen{q}\,\gen{\Sigma} = -\gen{q}\,.
\end{equation}

\subsection{One-particle (short) representations}
We are interested in representations satisfying the shortening condition
\begin{equation}
\label{eq:shortening}
    \gen{H}\,\widetilde{\gen{H}} = \gen{C}\,\bar{\gen{C}}\,.
\end{equation}
We will label the representations by the eigenvalues of the central charges $(H,\tilde{H},C,\bar{C})$. For unitary representations we will have
\begin{equation}
    H\geq0\,,\qquad
    \tilde{H}\geq0\,,\qquad
    C\,\bar{C}\geq0\,.
\end{equation}
Any such representation of the algebra~\eqref{eq:smallalgebra} is two-dimensional, consisting of one Boson and  one Fermion, and it takes the form
\begin{equation}
    \gen{q}\,\ket{\phi}=a\,\ket{\varphi}\,,\qquad
    \gen{s}\,\ket{\varphi}=a^*\,\ket{\phi}\,,\qquad
    \tilde{\gen{s}}\,\ket{\phi}=b^*\,\ket{\varphi}\,,\qquad
    \tilde{\gen{q}}\,\ket{\varphi}=b\,\ket{\phi}\,,
\end{equation}
where all the other actions of the generators are trivial. In our notation $\ket{\phi}$ is the highest-weight state of the representation (be that a Boson or a Fermion) and $\ket{\varphi}$ is the lowest-weight state (be that a Fermion or a Boson). Picking the basis ($(\ket{\phi},\ket{\varphi})$ we may explicitly write
\begin{equation}
\label{eq:representations}
    \gen{q}=\left(
    \begin{array}{cc}
         0&0  \\
         a&0 
    \end{array}
    \right)\,,\qquad
    \gen{s}=\left(
    \begin{array}{cc}
         0&a^*  \\
         0&0 
    \end{array}
    \right)\,,\qquad
    \tilde{\gen{s}}=\left(
    \begin{array}{cc}
         0&0  \\
         b^*&0 
    \end{array}
    \right)\,,\qquad
    \tilde{\gen{q}}=\left(
    \begin{array}{cc}
         0&b  \\
         0&0 
    \end{array}
    \right)\,.
\end{equation}
In terms of these representation coefficients the central charges are
\begin{equation}
    H=|a|^2\,,\qquad \tilde{H}=|b|^2\,,\qquad
    C=ab\,,\qquad \bar{C}=a^*b^*\,.
\end{equation}
Finally, it is convenient to introduce the combinations of the central charges
\begin{equation}
    \gen{E}=\gen{H}+\widetilde{\gen{H}}\,,\qquad
    \gen{M}=\gen{H}-\widetilde{\gen{H}}\,,
\end{equation}
and correspondingly we introduce their eigenvalues $E$ and $M$.

\subsection{Physical value of representation parameters}

For Ramond-Ramond backgrounds, the value of the charge $M$ is simply given by
\begin{equation}
    M\in\mathbb{Z}\,,
\end{equation}
where $M=0,\pm1$ correspond to the physical particles of the model and $|M|\geq2$ are the bound states. By semiclassical arguments it is possible to determine that, up to an overall phase that depends on the boundary conditions of the fields, it must be~\cite{Borsato:2014exa,Borsato:2014hja}
\begin{equation}
    \gen{C}\approx \frac{h}{2}(e^{+i\gen{p}}-1)\,,\qquad
    \bar{\gen{C}}\approx \frac{h}{2}(e^{-i\gen{p}}-1)\,,
\end{equation}
where $\gen{p}$ is the worldsheet momentum of the state (\textit{i.e.}, the generator of spatial translations on the worldsheet), and $h$ is a coupling constant related to the string tension.
For the one-particle representation hence we set
\begin{equation}
    C=+i\frac{h}{2}(e^{+ip}-1)e^{+2i\xi}\,,\qquad
    \bar{C}=-i\frac{h}{2}(e^{-ip}-1)e^{-2i\xi}\,,
\end{equation}
where $p$ is the momentum of the particle, and $\xi\in\mathbb{R}$ is a representation parameter related to the phase redundancy.
Using the shortening condition~\eqref{eq:shortening} we find the dispersion relation
\begin{equation}
    E(p,M)=\sqrt{M^2+4h^2\sin^2\Big(\frac{p}{2}\Big)}\,.
\end{equation}

For $M\geq 0$ we can define the representation parameters as  follows
\begin{equation}
    a=\eta_p e^{i\xi}\,,\qquad
    a^*=e^{-\tfrac{i}{2}p}\eta_p e^{-i\xi}\,,\qquad
    b=-e^{-\tfrac{i}{2}p}\frac{\eta_p}{x^-_p}e^{i\xi}\,,\qquad
    b^*=-\frac{\eta_p}{x^+_p}e^{-i\xi}\,.
\end{equation}
It is worth noting that it would be possible to redefine
\begin{equation}
\label{eq:automorphism}
    a\to e^{+i\alpha}\,a\,,\qquad
    b\to e^{-i\alpha}\,b\,,
\end{equation}
with~$\alpha$ a real parameter; this is an automorphism of the algebra. We have introduced the parameters
\begin{equation}
    \eta_p = e^{ip/4}\,\sqrt{\frac{ih}{2}(x^-_p - x^+_p)}\,,
\end{equation}
while the Zhukovsky variables satisfy
\begin{equation}
    x^+_p +\frac{1}{x^+_p} -x^-_p - \frac{1}{x^-_p} = \frac{2i}{h}\,|M|\,,\qquad
    \frac{x^+_p}{x^-_p}=e^{ip}\,,
\end{equation}
and may be parametrised as
\begin{equation}
\label{eq:Zhukovsky}
    x^\pm_p =e^{\pm i p/2}\,\frac{|M|+\sqrt{M^2+4h^2\sin^2(p/2)}}{2h\, \sin(p/2)}\,.
\end{equation}
We immediately find that the eigenvalue of $\gen{M}$ is given by $|a|^2-|b|^2=|M|\geq0$. We call these representations ``left'' if $M\geq1$ and ``massless'' if $M=0$. It should also be noted that we may also consider representations where the eigenvalue of $M$ is negative. Those are called ``right'' representations and are obtained by exchanging $a\leftrightarrow b$, see~\cite{Borsato:2013qpa}.%
\footnote{
It is possible to obtain the massless modes by setting $M=0$ in the left or in the right representations. The two procedures yield unitarily equivalent representations~\cite{Borsato:2014hja}.
}
Here we will focus on massless and ``left'' representations only, because those will be sufficient to make our points about the analytic structure of the scattering matrices involving massless modes. The discussion of the ``right'' representations would be the same.
In terms of the Zhukovsky variables the energy takes the simple form
\begin{equation}
    E(p,M)= \frac{h}{2i}\left(
    x^+_p-\frac{1}{x^+_p}-x^-_p+\frac{1}{x^-_p}
    \right)\,.
\end{equation}
Finally, we note that for the one-particle representation, we  shall fix
\begin{equation}
    \xi=0\,.
\end{equation}

Much like in $AdS_5\times S^5$, for any $|M|\geq1$ we could also uniformise this dispersion in terms of Jacobi elliptic functions on a torus whose elliptic modulus is related to~$h$, see \textit{e.g.}~\cite{Arutyunov:2009ga}.
The case of $M=0$ is quite special because the dispersion relation appears not to be analytic,
\begin{equation}
    E(p,0)=2h\,\left|\sin\Big(\frac{p}{2}\Big)\right|\,.
\end{equation}
Note however that the representation is $2\pi$-periodic. This is not immediately obvious from the form of the representation coefficients, but it is clear from the form of the central charges. Indeed, up to an automorphism of the form~\eqref{eq:automorphism}, the representation coefficients can be made $2\pi$-periodic. Hence, by picking as fundamental region $0\leq p <2\pi$, we relegate the non-analytic points at the boundary of the fundamental region. We shall see that this is actually not possible when considering the representations that appear in the S~matrix.

\subsection{Two-particle representation}

We can construct the two-particle representation by considering, for a generic supercharge~$\gen{q}$,
\begin{equation}
    \gen{q}(p_1,p_2) = \gen{q}(p_1;\xi_1)\otimes\gen{1} +\gen{\Sigma}\otimes\gen{q}(p_2;\xi_2)\,,
\end{equation}
where
\begin{equation}
\label{eq:xi12}
    \xi_1 =0\,,\qquad
    \xi_2=\frac{1}{2}p_1\,.
\end{equation}
It is possible to check that this is indeed a representation with
\begin{equation}
    M=M_1+M_2\,,\qquad  E= E(p_1,M_1)+E(p_2,M_2)\,,
\end{equation}
as well as
\begin{equation}
    C=\frac{ih}{2}(e^{i(p_1+p_2)}-1)\,,\qquad
    \bar{C}=C^*\,.
\end{equation}
To obtain the form of $C$ and $\bar C$, which is expected from string theory, it is crucial to impose~\eqref{eq:xi12}~\cite{Arutyunov:2006yd,Borsato:2014hja}. Equivalently, we could say that the algebra acting on the two-particle Hilbert space, or better yet on the Fock space as a whole, is a Hopf algebra with a non-trivial coproduct.

By using this representation, we can formulate invariance conditions on the S~matrix. Strictly speaking of course, we should be considering representations of the full algebra~\eqref{eq:bigalgebra}, but since those are induced by the representations of the smaller algebra~\eqref{eq:smallalgebra}, for the purpose of our discussion we may consider the latter.%
\footnote{This smaller algebra is also the one that dictates the form of the $AdS_3\times S^3\times S^3\times S^1$ S~matrix~\cite{Borsato:2012ud,Borsato:2012ss}, so that it is of interest in and of itself.} 
The condition for invariance of the S~matrix~$\gen{S}(p_1,p_2)$ is
\begin{equation}
\label{eq:SmatrixInvariance}
    \gen{S}(p_1,p_2)\,\gen{q}(p_1,p_2) = \gen{q}(p_2,p_1)\,\gen{S}(p_1,p_2)\,.
\end{equation}
Note that in this condition the S~matrix permutes the excitations scattered, \textit{i.e.}\  it acts for instance as
\begin{equation}
    \gen{S}\,\ket{\phi_1,\phi_2} = S_{0}(p_1,p_2)\,\ket{\phi_2,\phi_1}\,,
\end{equation}
for some coefficient~$S_{0}(p_1,p_2)$. The fact that for the example considered above the scattering is diagonal is a consequence of the fact that we are scattering two highest-weight states.
As it turns out, the condition~\eqref{eq:SmatrixInvariance} is sufficient to fix the S~matrix scattering two irreducible representations of the type we described up to an overall pre-factor, the so-called dressing factor. We will discuss the most important constraints on these dressing factors below.

\subsection{Momentum periodicity of the two-particle representation}
We now come to an important point concerning periodicity of the momenta $p_1$ and $p_2$ which was perhaps not appreciated in the existing literature. It is true that the central charges of the two-particle representation are invariant under
\begin{equation}
    p_1\to p_1 + 2\pi\,,\qquad\text{and/or}\qquad
    p_2\to p_2+2\pi\,.
\end{equation}
However, \emph{the S~matrix is not invariant under such a transformation}. In fact, the S~matrix is not even invariant under a shift of both momenta by $2\pi$. The reason lies in the non-trivial coproduct which we had to use to construct the two-particle representation. That coproduct depends on $e^{ip_1/2}=e^{i\xi_2}$ which is only $4\pi$-periodic. The issue of course does not concern only $p_1$, because in~\eqref{eq:SmatrixInvariance} we also need $\gen{q}(p_2,p_1)$, which involves~$e^{ip_2/2}$.

Based on our discussion of the one-particle representation, this may seem contradictory: if the central charges are $2\pi$-periodic, should it not be possible and perhaps advisable to find an automorphism that makes the two-particle representation coefficients $2\pi$-periodic too? This is what we argued one could do for the one-particle representation, after all.
While it is possible to find such an automorphism, it is clear from our discussion that it would not preserve the coproduct, hence yielding a fundamentally different Hopf algebra. In more physical terms. this would amount to a change of the two-particle basis which is not induced by a change of the one-particle basis. While the latter is merely a redefinition of the states, the former yields a genuinely different, albeit related, S~matrix.
Indeed this observation is not surprising at all in light of what happens for the $AdS_5\times S^5$ S~matrix, whose kinematics is very similar to the case at hand when $|M|\geq 1$. In fact, as detailed in ref.~\cite{Arutyunov:2009ga}, the matrix-part of the S~matrix has non-trivial monodromies under shifts of~$2\pi$.
In the case of  $AdS_5\times S^5$ however this is of little concern. In fact, the matrix part of the S~matrix is a meromorphic function on two copies of the rapidity torus. Even when including the dressing factor, no additional issue arises, as the dressing factor is meromorphic in a large region of the torus which includes a neighborhood of the real-momentum line~\cite{Arutyunov:2009kf}. Therefore, analytically continuing $p$ to any other real value such as $p+2\pi$ is straightforward. The same is also true for $AdS_3\times S^3\times T^4$ when $|M|\geq 1$.

This observation becomes much more troubling when we consider the case of $M=0$, \textit{i.e.}\ of massless particles. As emphasised, the dispersion relation \emph{is not analytic}. In fact, strictly speaking the dispersion splits into \emph{two distinct functions}, each analytic in its own domain,
\begin{equation}
\label{eq:chiiraldispersion}
E(p,0)=\begin{cases}
-2h\,\sin(p/2) & -2\pi\leq p\leq 0\qquad (\text{mod}4\pi)\,,\\
+2h\,\sin(p/2) & \quad0\leq p\leq 2\pi\qquad (\text{mod}4\pi)\,.
\end{cases}
\end{equation}
This is of course analogous to the chiral--anti-chiral split that we would encounter in massless relativistic theories with $E(p)=c|p|$. As a consequence, we cannot analytically continue the momentum from the region $-2\pi\leq p\leq 0$ to the region $0\leq p\leq 2\pi$ or viceversa.
Because of this impossibility, and because the two-particle representation is not $2\pi$-periodic, for the purposes of studying the S~matrix it is fundamentally different to consider $0\leq p\leq 2\pi$ or $-2\pi\leq p\leq 0$.

Having realised that the two regions appearing in~\eqref{eq:chiiraldispersion} are fundamentally different when it comes to the S~matrix, it is now natural to ask how we should define the fundamental region for the S~matrix. At the very least, the physical region should include a neighborhood of $p=0$. In fact, the only known perturbative definition of the S~matrix comes from the lightcone gauge-fixed string non-linear sigma model. In particular, the near-BMN expansion is a large-$h$ and small-$p$ expansion, which ``zooms in'' around $p=0$, where the dispersion is relativistic at leading order. Since here $p\lesssim 0$ is not equivalent to $p\lesssim2\pi$, we need to consider both of the regions appearing in eq.~\eqref{eq:chiiraldispersion}. Strictly speaking, we have distinct S~matrices depending on the value of the momentum. This is analogous to the relativistic case. In that case, when massless particles are present, the branch points of the  Mandelstam plane collide and the two branches become separate, see~\cite{Polyakov:1983tt,Zamolodchikov:1992zr,Fendley:1993wq} as well as Appendix~A of~\cite{Borsato:2016xns} for a discussion of this point.
Unfortunately, much of the discussion of the massless S~matrix for $AdS_3\times S^3\times T^4$ was so far performed in the region~$0\leq p\leq 2\pi$ only~\cite{Bombardelli:2018jkj,Fontanella:2019baq,Fontanella:2019ury}. In the rest of the paper we shall confirm that considering both regions is necessary to reproduce some desirable physical properties of the S~matrix.

\section{Chiral and antichiral particles}
\label{sec:chiralsplit}
We have concluded in the previous section that the S~matrix is only $4\pi$-periodic in either momentum variable, but it is not $2\pi$-periodic. This is particularly relevant when considering massless particles, which have a non-analytic dispersion relation $E(p)=2h|\sin(p/2)|$. We will make this explicit by adopting a parametrisation due to Fontanella and Torrielli~\cite{Fontanella:2019baq}.

\subsection{Fundamental domain}
As we have discussed, while the dispersion relation is $2\pi$-periodic, we expect that the two-particle representation is not. Instead, it will be $4\pi$-periodic. Given this, we should decide what is the fundamental domain for the momentum~$p$, \textit{i.e.}\ the interval on the real line where~$p$ takes values for  physical particles with real energy (as opposed to complex ones, such as the one that one may encounter in bound states). Several nonequivalent choices come naturally to mind: the two asymmetric choices $0\leq p\leq 2\pi$, $-2\pi\leq p\leq0$, the symmetric choice $-\pi\leq p\leq\pi$ and the maximal choice $-2\pi\leq p\leq 2\pi$.

There are several arguments that we may try to employ to decide this, including analogy with the case of (massive) particles in $AdS_5\times S^5$ which would suggest as a natural choice $-\pi\leq p\leq\pi$. A rather physical observation is that we want a picture that is consistent with the near-BMN limit. In this limit, $p$ is rescaled so that we ``zoom in'' around zero, which excludes the asymmetrical choices. To distinguish between $-\pi\leq p\leq\pi$ and the maximal choice $-2\pi\leq p\leq 2\pi$ we can consider some other property of the model.
In integrable models we expect the Bethe equations (as well as the full spectrum) to enjoy some special symmetry related to the existence of symmetry descendants and/or protected states. We expect that, given a solution of the Bethe equations (\textit{i.e.}, a physical state of the theory), we should be able to construct another state which is obtained from the previous by acting with a (super)symmetry generator. Due to the specific form of the supersymmetry algebra, creating such a descendant must change the energy of a state by a fixed amount, independent from the original state and the value  of the coupling constant~$h$. One necessary condition for this to be the case is that the new state is obtained from the old one by adding one particle of momentum $p_*$, where $p_*$ is such that
\begin{equation}
    \frac{\partial}{\partial h}E(p_*,M)=0\,.
\end{equation}
Clearly this happens if $p_*=0$ mod$2\pi$. Indeed, for such a choice $E(p_*,M)=|M|$.
In the case of massive particles with $M=\pm1$, descendants should correspond to $p_*=0$, and to avoid overcounting we should take $-\pi<p<\pi$~\cite{Borsato:2013qpa}. Already this may be reason enough to pick $-\pi<p<\pi$ as a fundamental domain (as it would be strange to have different domains for massive and massless particles). However, we can also see this more directly.

For the particular case of massless excitations, the energy actually vanishes and the interpretation of $p_*=0$ is slightly different: such zero-energy modes are necessary to reproduce the degeneracy of the spectrum~\cite{Borsato:2016kbm,Baggio:2017kza}. In particular, they account for the spectrum of half-BPS multiplets and for the existence of flat directions along~$T^4$ in $AdS_3\times S^3 \times T^4$.%
\footnote{This is different from $AdS_5\times S^5$, where for a given choice of the lightcone gauge-fixing there is only one half-BPS state, which is the trivial vacuum. Note also that adding zero-energy excitations  in $AdS_3\times S^3 \times T^4$ does not correspond to acting with a $\gen{psu}(1,1|2)^{\oplus 2}$ generator, see~\cite{Borsato:2016kbm,Baggio:2017kza,Majumder:2021zkr}.} 
At the very least, we have that both $p_*=0^+$ and $p_*=0^-$ should be among the possible solutions of the Bethe equations. As discussed, this is because we want this picture to persist in the near-BMN limit. It turns out that the states at zero momentum are precisely what is needed to account for the expected degeneracy of the model~\cite{Borsato:2016kbm,Baggio:2017kza,Majumder:2021zkr}.%
\footnote{See also~\cite{Dei:2018mfl} for a detailed discussion of the counting of states, albeit in a slightly different setup.}
Hence we may rule out $-2\pi\leq p\leq 2\pi$ (which would give too many zero-energy states), and we conclude that the physical region for real particles should be 
\begin{equation}
    -\pi\leq p\leq \pi\qquad\text{(fundamental domain).}
\end{equation}
We will see later on an \textit{a posteriori} justification of this choice, when we will discuss the explicit form of the S~matrix.

\subsection{Rapidity parametrisation}
Let us start by observing that for massless particle the Zhukovsky variables may be defined by setting $M=0$ in~\eqref{eq:Zhukovsky}
\begin{equation}
    x^\pm_p = \frac{\sqrt{\sin^2(p/2)}}{\sin(p/2)}\,e^{\pm i p/2}\,,
\end{equation}
so that
\begin{equation}
    x^+_p\,x^-_p = 1\,.
\end{equation}
Note that with this definition $x_p^+$ lies on the upper-half circle for real momentum.
As a result, we can work in terms of a single Zhukovsky variable $x_p$
\begin{equation}
    x_p = x^+_p\big|_{M=0}\,.
\end{equation}
It turns out that it is very convenient to introduce a relativistic rapidity $\gamma_p$ such that~\cite{Fontanella:2019baq}
\begin{equation}
    x_p= \frac{i-e^{\gamma_p}}{i+e^{\gamma_p}}\,.
\end{equation}
Real values of the momentum correspond to real~$\gamma_p$.
In these terms, the energy is manifestly positive for real momentum, and it reads
\begin{equation}
    E(\gamma_p,0)= \frac{2h}{\cosh\gamma_p}\,.
\end{equation}
In order to parametrise the representation, we have to make explicit the relationship between $\gamma_p$ and $p$ itself. This needs to be done in the whole region $-\pi \leq p \leq \pi$. 
One possibility is to parametrise $\gamma_p$ so that $p(\gamma)$ has the correct range by setting
\begin{equation}
    p=-i\log\left(\frac{i-e^{\gamma}}{i+e^{\gamma}}\right)^2\,,\qquad \text{for}\quad \gamma\in\mathbb{R}\,,\quad -\pi\leq p\leq\pi\,,
\end{equation}
with the standard choice of branch  for the logarithm. However, this choice makes the function $e^{ip/2}$ (which was the functions spoiling $2\pi$-periodicity) non-analytic. We find it convenient to resolve this non-analiticity by explicitly employing two distinct parametrisations depending on the sign of (the real part of) $p$ and $\gamma$. Namely, we set firstly
\begin{equation}
    \gamma_p =\log\left(-\cot\frac{p}{4}\right)\,,\quad
    p=-2i\log\left(-\frac{i-e^{\gamma_p}}{i+e^{\gamma_p}}\right)\,,\qquad
    \text{for}\quad\gamma_p\geq0\,,\quad -\pi\leq p \leq 0\,.
\end{equation}
Notice that we can analytically continue $\gamma$ wherever we want, including to the negative half-line, but this would not give us the $0\leq p \leq \pi$ region. In fact, the negative half-line corresponds to the (unphysical) region $[-2\pi,-\pi]$. We could in principle cross the cut of the logarithm, but this would only produce a monodromy which is a multiple of $4\pi$, which would take us to other unphysical regions. To describe $0\leq p \leq \pi$ we need instead
\begin{equation}
    \gamma_p =\log\left(\tan\frac{p}{4}\right)\,,\quad
    p=-2i\log\left(+\frac{i-e^{\gamma_p}}{i+e^{\gamma_p}}\right)\,,\qquad
    \text{for}\quad \gamma_p\leq0\,,\quad 0\leq p \leq \pi\,.
\end{equation}
With these definitions we can check that the function $e^{ip/2}$  has the following representation
\begin{equation}
    e^{ip/2}=\begin{cases}
    -x=-\displaystyle\frac{i-e^{\gamma_p}}{i+e^{\gamma_p}}\qquad&\gamma_p\geq0\,,\quad-\pi\leq p\leq 0\,,\\[0.4cm]
    +x=+\displaystyle\frac{i-e^{\gamma_p}}{i+e^{\gamma_p}}\qquad&\gamma_p\leq0\,,\quad\quad0\leq p\leq \pi\,,
    \end{cases}
\end{equation}
as desired.  We conclude that we have to treat separately the two cases
\begin{equation}
\begin{aligned}
    -\pi\leq p\leq 0&\qquad\text{``antichiral''}\qquad&&\text{denoted by ``$-$''},\\
    0\leq p\leq \pi&\qquad\text{``chiral''}\qquad&&\text{denoted by ``$+$''}.
\end{aligned}
\end{equation}

\subsection{Massless representation, one particle}
Let us now rewrite the representation coefficients in terms of the rapidity~$\gamma_p$. We will distinguish the chiral and antichiral case. In both cases we find a rather simple form for the representation coefficients.

\paragraph{Antichiral case}
\begin{equation}
    a_{-}=b_-=\sqrt{2h}\frac{e^{\gamma/2}}{e^{\gamma}+i}\,,\qquad
    a_{-}^*=b_-^*=\sqrt{2h}\frac{e^{\gamma/2}}{e^{\gamma}-i}\qquad
    (\gamma\geq0,\ -\pi\leq p\leq 0).
\end{equation}

\paragraph{Chiral case}
\begin{equation}
    a_{+}=-b_+=i\sqrt{2h}\frac{e^{\gamma/2}}{e^{\gamma}+i}\,,\qquad
    a_{+}^*=-b_+^*=-i\sqrt{2h}\frac{e^{\gamma/2}}{e^{\gamma}-i}\qquad
    (\gamma\leq0,\ 0\leq p\leq \pi).
\end{equation}
It is immediately clear that this case cannot be obtained from the previous one by analytic continuation in $\gamma$ since in one case $a_-=b_-$ while in the other $a_+=-b_+$.

\paragraph{Automorphism.}
We can easily see that we can relate the two one-particle representations by means of an automorphism of the type~\eqref{eq:automorphism}, namely
\begin{equation}
    a_-\to +i\,a_- = a_+\,,\qquad
    b_-\to -i\,b_-=b_+\,.
\end{equation}
This is well in accord with our observation that, when considering the one-particle representation only, $2\pi$-periodicity is guaranteed up to an isomorphism or a change of basis.

\paragraph{Matrix representation and change of basis.}
To make the above automorphism more explicit, let us introduce the basis $(\ket{\phi},\ket{\varphi})$ where $\ket{\phi}$ is the highest-weight state (be it a Boson or a Fermion) and $\ket{\varphi}$ is the lowest-weight one. The supercharges can be represented as $2\times2$ matrices,
\begin{equation}
    \gen{q}_\pm=\left(
    \begin{array}{cc}
         0&0  \\
         a_\pm&0 
    \end{array}
    \right)\,,\qquad
    \gen{s}_\pm=\left(
    \begin{array}{cc}
         0&a^*_\pm  \\
         0&0 
    \end{array}
    \right)\,,\qquad
    \tilde{\gen{s}}_\pm=\left(
    \begin{array}{cc}
         0&0  \\
         b^*_\pm&0 
    \end{array}
    \right)\,,\qquad
    \tilde{\gen{q}}_\pm=\left(
    \begin{array}{cc}
         0&b_\pm  \\
         0&0 
    \end{array}
    \right)\,,
\end{equation}
where the indices $\pm$ are just to remind us that we have two different representations for chiral and antichiral particles.
Introducing the diagonal change-of-basis $\gen{v}$
\begin{equation}
    \gen{v}=\left(
    \begin{array}{cc}
         e^{+i\pi/4}&0  \\
         0&e^{-i\pi/4} 
    \end{array}
    \right)\,,
\end{equation}
we find 
\begin{equation}
\label{eq:oneparticleisomorphism}
    \gen{v}^\dagger\gen{q}_-\gen{v}=\gen{q}_+\,,\qquad
    \gen{v}^\dagger\gen{s}_-\gen{v}=\gen{s}_+\,,\qquad
    \gen{v}^\dagger\tilde{\gen{s}}_-\gen{v}=\tilde{\gen{s}}_+\,,\qquad
    \gen{v}^\dagger\tilde{\gen{q}}_-\gen{v}=\tilde{\gen{q}}_+\,.
\end{equation}

\subsection{Two-particle representation}
When constructing the two-particle representations we should distinguish several cases. In fact, we may have one or both massless particles. Let us start with the case where one particle is massless and the other is massive. For definiteness, we take the massive particle to be in the ``left'' representation, \textit{i.e.}\ to have~$M\geq1$. We will denote the supercharges in the massive representation by $\gen{q}_\bullet$.

\paragraph{One particle is antichiral, the other massive.}
In this case we write, if the first particle is massless,
\begin{equation}
    \gen{q}_{-\bullet}(\gamma,p) = \gen{q}_-(\gamma)\otimes\gen{1}- \frac{i-e^{\gamma}}{i+e^{\gamma}}\,\gen{\Sigma}\otimes\gen{q}_\bullet(p)\,,
\end{equation}
where we made the coproduct explicit. Related to this, we have also
\begin{equation}
\label{eq:secondmixed}
     \gen{q}_{\bullet-}(p,\gamma) = \gen{q}_\bullet(p)\otimes\gen{1}+ e^{i p/2}\,\gen{\Sigma}\otimes\gen{q}_{-}(\gamma)\,.   
\end{equation}
These are the two expressions that are needed to determine the S~matrix~\eqref{eq:SmatrixInvariance}.

\paragraph{One particle is chiral, the other massive.}
Now we write, if the first particle is massless,
\begin{equation}
    \gen{q}_{+\bullet}(\gamma,p) = \gen{q}_+(\gamma)\otimes\gen{1}+ \frac{i-e^{\gamma}}{i+e^{\gamma}}\,\gen{\Sigma}\otimes\gen{q}_\bullet(p)\,,
\end{equation}
and we see now that the coefficient of the second terms has the opposite sign. Related to this, we have also
\begin{equation}
     \gen{q}_{\bullet+}(p,\gamma) = \gen{q}_\bullet(p)\otimes\gen{1}+ e^{i p/2}\,\gen{\Sigma}\otimes\gen{q}_{+}(\gamma)\,.   
\end{equation}
Note that here the coefficients have the same signs as in~\eqref{eq:secondmixed}.

\paragraph{General formula for the mixed-mass case.}
It is convenient to write compactly the cases above as
\begin{equation}
\begin{aligned}
        \gen{q}_{\pm\bullet}(\gamma,p) &= \gen{q}_\pm(\gamma)\otimes\gen{1}\pm \frac{i-e^{\gamma}}{i+e^{\gamma}} \,\gen{\Sigma}\otimes\gen{q}_\bullet(p)\,,\\
        \gen{q}_{\bullet\pm}(p,\gamma) &= \gen{q}_\bullet(p)\otimes\gen{1}+ e^{i p/2}\,\gen{\Sigma}\otimes\gen{q}_{\pm}(\gamma)\,.   
\end{aligned}
\end{equation}

\paragraph{Both particles are massless.} Building on the two cases above, it is not hard to write a compact formula for the case where both particles are massless:
\begin{equation}
        \gen{q}_{\pm\alpha}(\gamma_1,\gamma_2) = \gen{q}_\pm(\gamma_1)\otimes\gen{1}\pm
        \frac{i-e^{\gamma_1}}{i+e^{\gamma_1}} \,\gen{\Sigma}\otimes\gen{q}_\alpha(\gamma_2)\,,\qquad
        \alpha\in\{+,-\}.
\end{equation}

\paragraph{Relation between the various representations.}
Recalling the relation for the one-particle representation~\eqref{eq:oneparticleisomorphism} as well as the identity~\eqref{eq:sigmaidentity} for the Fermion sign matrix, we can relate various representations considered above by a change of the two-particle basis.
We have
\begin{equation}
\begin{aligned}
    \gen{q}_{+\bullet}(\gamma,p) &=\gen{v}^\dagger\otimes\gen{\Sigma}\cdot
    \gen{q}_{-\bullet}(\gamma,p)\cdot \gen{v}\otimes\gen{\Sigma}\,,\\
    \gen{q}_{\bullet+}(p,\gamma) &= \ \gen{1}\otimes\gen{v}^\dagger\cdot\gen{q}_{\bullet-}(p,\gamma)\ \cdot\gen{1}\otimes\gen{v}\,,
\end{aligned}
\end{equation}
as well as
\begin{equation}
\begin{aligned}
    \gen{q}_{+\alpha}(\gamma_1,\gamma_2) &=\gen{v}^\dagger\otimes\gen{\Sigma}\cdot
    \gen{q}_{-\alpha}(\gamma_1,\gamma_2)\cdot \gen{v}\otimes\gen{\Sigma}\,,\\
    \gen{q}_{\alpha+}(\gamma_1,\gamma_2) &=\ \gen{1}\otimes\gen{v}^\dagger\cdot
    \gen{q}_{\alpha-}(\gamma_1,\gamma_2)\cdot\ \gen{1}\otimes\gen{v}\,,
\end{aligned}
\end{equation}
and
\begin{equation}
    \gen{q}_{++}(\gamma_1,\gamma_2) =\gen{v}^\dagger\otimes(\gen{\Sigma}\gen{v}^\dagger)\cdot
    \gen{q}_{--}(\gamma_1,\gamma_2)\cdot \gen{v}\otimes(\gen{v}\gen{\Sigma})\,.
\end{equation}

\section{S matrices}
\label{sec:Smatrices}
We can define several S~matrices for the various branches of the massless excitations by specialising eq.~\eqref{eq:SmatrixInvariance} to the various cases. We have
\begin{equation}
\begin{aligned}
    \gen{S}_{\alpha\bullet}(\gamma,p)\,\gen{q}_{\alpha\bullet}(\gamma,p) &= \gen{q}_{\bullet\alpha}(p,\gamma)\,\gen{S}_{\alpha\bullet}(\gamma,p)\,,\\
    \gen{S}_{\bullet\alpha}(p,\gamma)\,\gen{q}_{\bullet\alpha}(p,\gamma) &= \gen{q}_{\alpha\bullet}(\gamma,p)\,\gen{S}_{\bullet\alpha}(p,\gamma)\,,\\
\end{aligned}
\end{equation}
and
\begin{equation}
        \gen{S}_{\alpha\beta}(\gamma_1,\gamma_2)\,\gen{q}_{\alpha\beta}(\gamma_1,\gamma_2) = \gen{q}_{\beta\alpha}(\gamma_2,\gamma_1)\,\gen{S}_{\alpha\beta}(\gamma_1,\gamma_2)\,.
\end{equation}
These equations are sufficient to define all of the S~matrices up to their dressing factors.

\subsection{Identities between the various S~matrices.}
The fact that the various representations introduced above are related by a change of the two-particle basis means that so are the S~matrices. Let us consider, for instance, the invariance condition
\begin{equation}
\gen{S}_{+\bullet}(\gamma,p)\,\gen{q}_{+\bullet}(\gamma,p) = \gen{q}_{\bullet+}(p,\gamma)\,\gen{S}_{+\bullet}(\gamma,p)
\end{equation}
and use the relation between $\gen{q}_{+\bullet}(\gamma,p)$ and $\gen{q}_{-\bullet}(\gamma,p)$, as well as the one between $\gen{q}_{\bullet+}(p,\gamma)$ and $\gen{q}_{\bullet-}(p,\gamma)$. We get
\begin{equation}
    \gen{S}_{+\bullet}(\gamma,p)\left(
    \gen{v}^\dagger\otimes\gen{\Sigma}\cdot
    \gen{q}_{-\alpha}(\gamma_1,\gamma_2)\cdot \gen{v}\otimes\gen{\Sigma}
    \right)=\left(\gen{1}\otimes\gen{v}^\dagger\cdot
    \gen{q}_{\alpha-}(\gamma_1,\gamma_2)\cdot\gen{1}\otimes\gen{v}\right)\gen{S}_{+\bullet}(\gamma,p)\,,
\end{equation}
which we can recast as
\begin{equation}
\begin{aligned}
    \left(\gen{1}\otimes\gen{v}\cdot\gen{S}_{+\bullet}(\gamma,p)\cdot\gen{v}^\dagger\otimes \Sigma\right)
    &\gen{q}_{-\alpha}(\gamma_1,\gamma_2)\\
    &=\gen{q}_{\alpha-}(\gamma_1,\gamma_2)
    \left(\gen{1}\otimes\gen{v}\cdot\gen{S}_{+\bullet}(\gamma,p)\cdot\gen{v}^\dagger\otimes \Sigma\right)\,.
\end{aligned}
\end{equation}
This is the same invariance condition as for $\gen{S}_{-\bullet}(\gamma,p)$. We conclude that, up to possibly  an overall prefactor,
\begin{equation}
\begin{aligned}
    \gen{S}_{-\bullet}(\gamma,p)\approx \gen{1}\otimes\gen{v}\ \cdot\gen{S}_{+\bullet}(\gamma,p)\cdot\gen{v}^\dagger\otimes \Sigma\,,\\
    \gen{S}_{+\bullet}(\gamma,p)\approx \gen{1}\otimes\gen{v}^\dagger\cdot\gen{S}_{-\bullet}(\gamma,p)\cdot\gen{v}\ \otimes \Sigma\,.
\end{aligned}
\end{equation}
In a similar way we may derive
\begin{equation}
\begin{aligned}
    \gen{S}_{\bullet-}(p,\gamma)\approx \gen{v}\ \otimes\Sigma\cdot\gen{S}_{\bullet+}(p,\gamma)\cdot\gen{1}\otimes\gen{v}^\dagger,\\
    \gen{S}_{\bullet+}(p,\gamma)\approx \gen{v}^\dagger\otimes\Sigma\cdot\gen{S}_{\bullet-}(p,\gamma)\cdot\gen{1}\otimes\gen{v}\ .
\end{aligned}
\end{equation}
The equations for massless-massless scattering take very similar forms,
\begin{equation}
\begin{aligned}
    \gen{S}_{-\alpha}(\gamma_1,\gamma_2)\approx \gen{1}\otimes\gen{v}\ \cdot\gen{S}_{+\alpha}(\gamma_1,\gamma_2)\cdot\gen{v}^\dagger\otimes \Sigma\,,\\
    \gen{S}_{+\alpha}(\gamma_1,\gamma_2)\approx \gen{1}\otimes\gen{v}^\dagger\cdot\gen{S}_{-\alpha}(\gamma_1,\gamma_2)\cdot\gen{v}\ \otimes \Sigma\,,\\
    \gen{S}_{\alpha-}(\gamma_1,\gamma_2)\approx \gen{v}\ \otimes\Sigma\cdot\gen{S}_{\alpha+}(\gamma_1,\gamma_2)\cdot\gen{1}\otimes\gen{v}^\dagger,\\
    \gen{S}_{\alpha+}(\gamma_1,\gamma_2)\approx \gen{v}^\dagger\otimes\Sigma\cdot\gen{S}_{\alpha-}(\gamma_1,\gamma_2)\cdot\gen{1}\otimes\gen{v}\ .
\end{aligned}
\end{equation}
Building on these formulae we can also find
\begin{equation}
\begin{aligned}
    \gen{S}_{--}(\gamma_1,\gamma_2)\approx\gen{v}\ \otimes \Sigma\gen{v}\ \cdot\gen{S}_{++}(\gamma_1,\gamma_2)\cdot  \gen{v}^\dagger\otimes\Sigma\gen{v}^\dagger,\\
    \gen{S}_{-+}(\gamma_1,\gamma_2)\approx\gen{v}^\dagger\otimes \Sigma\gen{v}\ \cdot\gen{S}_{+-}(\gamma_1,\gamma_2)\cdot  \gen{v}^\dagger\otimes \Sigma\gen{v}\ .
\end{aligned}
\end{equation}
Despite the somewhat more symmetric form, these transformations are not induced by one-particle changes of basis (due to the Fermion-sign matrix $\Sigma$) so that the S~matrix undergoes a non-trivial monodromy.

\subsection{Explicit form of the S~matrices}
We should stress that this discussion of the S~matrices is perfectly compatible with that of refs.~\cite{Borsato:2014exa, Borsato:2014hja}, see also~\cite{Eden:2021xhe} for a concise summary of the results. In those references, the expressions for the S~matrix involve only the Zhukovsky parameters~$x^\pm_p$ and $x_p$ which are valid for any value of~$p$. The trouble only comes when we try to express the S~matrix in terms of~$p$ itself, or of the rapidity $\gamma_p$, because as we have seen the relation between $x_p$ and $p$, and between $\gamma_p$ and $p$, is not analytic. Therefore, the formulae~\cite{Borsato:2014exa, Borsato:2014hja}  can give us the correct expressions in terms of $\gamma_p$ as long as we are careful in distinguishing the chiral and antichiral regions. It is especially worth considering the massless-massless S~matrix because in that case it  takes a particularly simple form when expressed in terms of the rapidities~$\gamma_p$: in fact, it is of difference form~\cite{Fontanella:2019baq}.
A point which was not always emphasised in the literature~\cite{Bombardelli:2018jkj,Fontanella:2019ury} is that we are dealing in fact with four distinct S~matrices, each of which is of difference form. To make this transparent, let us write down the massless-massless S~matrix explicitly. Picking the basis $(\ket{\phi},\ket{\varphi})$ for both representations, and assuming for definiteness that in both cases the highest-weight state is a Boson, we have
\begin{equation}
\gen{S}_{\pm\pm}(\gamma_1,\gamma_2)=
\Sigma_{\pm\pm}(\gamma_1,\gamma_2)
\begin{pmatrix}
1   &    0   &   0 &  0 \\
0   &    \sech\frac{\gamma_{12}}{2}   &   \pm\tanh\frac{\gamma_{12}}{2} &  0 \\
0   &    \mp\tanh\frac{\gamma_{12}}{2}   &   \sech\frac{\gamma_{12}}{2} &  0 \\
0   &    0   &   0 &  1 
\end{pmatrix}\,,
\end{equation}
where $\gamma_{12}=\gamma_1 - \gamma_2$, while
\begin{equation}
\gen{S}_{\pm\mp}(\gamma_1,\gamma_2) =
\Sigma_{\pm\mp}(\gamma_1,\gamma_2) 
\begin{pmatrix}
1   &    0   &   0 &  0 \\
0   &    \mp i\,\sech\frac{\gamma_{12}}{2}   &   \mp\tanh\frac{\gamma_{12}}{2} &  0 \\
0   &    \mp\tanh\frac{\gamma_{12}}{2}   &   \mp i\,\sech\frac{\gamma_{12}}{2} &  0 \\
0   &    0   &   0 &  -1 
\end{pmatrix}
\,.
\end{equation}
Both formulae hold up to a dressing factor~$\Sigma_{\alpha\beta}(\gamma_1,\gamma_2)$.%
\footnote{While the matrix part of the massless-massless S~matrices depend only on $\gamma_1-\gamma_2$ as a consequence of the symmetries of the model, we will assume that, in full generality, the dressing factors may depend on $(\gamma_1,\gamma_2)$ separately. 
In fact, the dressing factor ``knows'' about the scattering of all particles of the model (which appear in the loops) including the massive excitations whose kinematics is not of difference form, and which we argue~\cite{Frolov:2021fmj} do contribute to the dressing factor. Instead, an hypothetical integrable model consisting only of massless particles would have a dressing factor of difference form.}
While all four matrices are of difference form, except maybe the dressing factors, they differ precisely by the monodromies discussed above.
Similar expressions can be written down for the mixed-mass S~matrices, though they are a little more bulky because, in that case, the result is not of difference form. We report them in appendix~\ref{app:smatrix}.

\subsection{Properties of the S~matrices}
We discuss here some important properties of the massless S~matrices, with particular emphasis on the properties more directly affected by the chiral--anti-chiral split.

\paragraph{Physical regions.}
Firstly, it is worth clarifying in what regions, for real particles (\textit{i.e.}, for particles having real energy), the scattering is physical. In principle one could ask the same question for particles of complex momentum too. In that case the discussion is much more subtle, and we postpone it to a future study~\cite{Frolov:2021fmj}. For particles of momentum $p_1$ and $p_2$ and mass $M_1,M_2$ we require that $S(p_1,p_2)$ is physical when
\begin{equation}
\label{eq:physicalmomenta}
    v_1> v_2\,,\qquad v_j = \frac{\partial}{\partial p}E(p, M_j)\Big|_{p=p_j}\,,
\end{equation}
where $v_1$ and $v_2$ are the group velocities of the two particles. This means that (in the ``in'' state) the ``leftmost'' particle  should go faster than the ``rightmost'' particle --- so that it can actually catch up with it and scatter. As a check, in the relativistic case where $E^2=M^2+p^2=M^2\cosh^2\theta$, we have that the Mandelstam $s$-variable can be expressed as
\begin{equation}
 s=M_1^2+M_2^2+2M_1M_2\cosh(\theta_1-\theta_2)\qquad\text{(relativistic case)}\,.   
\end{equation}
Then $v_j=\tanh\theta_j$ and clearly region where the scattering is physical is $\theta_1-\theta_2>0$ as expected.
With the condition \eqref{eq:physicalmomenta} we have that $S_{+-}(\gamma_1,\gamma_2)$ is always physical, while $S_{-+}(\gamma_1,\gamma_2)$ is never physical. However, in contrast with what we might expect in massless relativistic scattering, here $S_{\pm\pm}(\gamma_1,\gamma_2)$ may be physical, depending on the value of the rapidities. Similarly, $S_{\pm\bullet}(\gamma,p)$ and $S_{\bullet\pm}(p,\gamma)$ may result in physical scattering processes for suitable values of the velocities.
Let us emphasise that, while in the relativistic case $v_1-v_2>0$ is related by a boost to the more restrictive condition $(v_1>0)\wedge (v_2<0)$, this is not true in the non-relativistic case at hand. The condition $(v_1>0)\wedge (v_2<0)$ would force us to restrict to $S_{-+}(\gamma_1,\gamma_2)$ which does not appear sensible when we want to consider multi-particle scattering --- in that case, at least one pair of particles will have the same direction. Additionally, as we shall see below, $S_{-+}(\gamma_1,\gamma_2)$ by itself does not satisfy the Yang-Baxter equation, so that it is absolutely crucial to consider all of the various S~matrices introduced above.

\paragraph{Yang-Baxter equation.}
The mixed-mass and massless-massless S~matrix satisfy the Yang-Baxter equation~\cite{Borsato:2014exa,Borsato:2014hja}. It is worth writing down explicitly the relation for different chiralities. Let us start from the case where all particle have the same chirality, say ``plus''; then the relation takes the familiar form, 
\begin{equation}
\begin{aligned}
    &\gen{S}_{++}(\gamma_{2},\gamma_{3})\otimes\gen{1}\ \cdot\ 
    \gen{1}\otimes\gen{S}_{++}(\gamma_{1},\gamma_{3})\ \cdot\ 
    \gen{S}_{++}(\gamma_{1},\gamma_{2})\otimes\gen{1}\\
    &\qquad\qquad\qquad\qquad=
    \gen{1}\otimes\gen{S}_{++}(\gamma_{1},\gamma_{2})\ \cdot\ 
    \gen{S}_{++}(\gamma_{1},\gamma_{3})\otimes\gen{1}\ \cdot\ 
    \gen{1}\otimes\gen{S}_{++}(\gamma_{2},\gamma_{3})\,.
\end{aligned}
\end{equation}
A little more care is needed when one of the particles has a different chirality from the other two. Then we have \textit{e.g.}%
\footnote{%
This expression is interesting because it relates the scattering of same-chirality particles to that of opposite-chirality ones. It is intuitive that, as we will see below, neither unitarity nor parity may be used to relate $\gen{S}_{\pm\pm}(\gamma)$ to $\gen{S}_{\pm\mp}(\gamma)$. 
}
\begin{equation}
\label{eq:mixedYBE}
\begin{aligned}
    &\gen{S}_{+-}(\gamma_{2},\gamma_{3})\otimes\gen{1}\ \cdot\ 
    \gen{1}\otimes\gen{S}_{+-}(\gamma_{1},\gamma_{3})\ \cdot\ 
    \gen{S}_{++}(\gamma_{1},\gamma_{2})\otimes\gen{1}\\
    &\qquad\qquad\qquad\qquad=
    \gen{1}\otimes\gen{S}_{++}(\gamma_{1},\gamma_{2})\ \cdot\ 
    \gen{S}_{+-}(\gamma_{1},\gamma_{3})\otimes\gen{1}\ \cdot\ 
    \gen{1}\otimes\gen{S}_{+-}(\gamma_{2},\gamma_{3})\,.
\end{aligned}
\end{equation}
One important consequences is that it is not self-consistent to consider only $\gen{S}_{+-}(\gamma_1,\gamma_2)$. In a sense, $\gen{S}_{+-}(\gamma_1,\gamma_2)$ is the most physical block of the S~matrix, because it is the one that appears when the momenta are small, \textit{i.e.}\ has in the near-BMN perturbation theory. We might hence be tempted to define the S~matrix in terms of  $\gen{S}_{+-}(\gamma_1,\gamma_2)$ only, and set it to be trivial in the same-chirality sector. Equation \eqref{eq:mixedYBE} shows that such a definition would violate the Yang-Baxter equation. In fact, it is easy to see that once  $\gen{S}_{+-}(\gamma_1,\gamma_2)$ is given, the solution for  $\gen{S}_{++}(\gamma_1,\gamma_2)$ is unique up to a dressing factor as \eqref{eq:mixedYBE} is a linear equation. We conclude that in the Bethe-Yang equation it will be necessary to deal with all the various S~matrices $\gen{S}_{\alpha\beta}$. The same reasoning holds for mixed-mass scattering.

\paragraph{Zero-momentum particles.}
We expect the S~matrices to simplify substantially when either momentum goes to zero. Recall that, perhaps counter-intuitively, when the momentum of massless particle goes to zero the velocity (or more precisely, the absolute value of the velocity) takes its maximum value. Recalling that $v=\partial_p H$, for a massless particle we find $|v(p)|=h\cos(p)$ for $-\pi\leq p\leq \pi$, and in particular
\begin{equation}
\begin{aligned}
    p\to 0^+\quad \Leftrightarrow\quad v\to +h\quad  \Leftrightarrow\quad \gamma\to -\infty\,,\\
    p\to 0^-\quad \Leftrightarrow\quad v\to -h\quad  \Leftrightarrow\quad \gamma\to +\infty\,,
\end{aligned}
\end{equation}
Moreover, a direct computation shows that for a massive particle with $|M|\geq 1$ we always have $|v(p,M)|<h$. This is important in light of our identification of the physical region~\eqref{eq:physicalmomenta}. In a physical scattering, involving massless particles, we can take the momentum of the first (massless) particle $p_1\to0^+$, and/or we can take the momentum of the second (massless) particle $p_2\to0^-$, but not viceversa. In either case, for massless-massless scattering, this always (and quite consistently) gives $\gamma_{12}\to-\infty$. When we encounter a zero-momentum particle therefore we need to consider
\begin{equation}
\label{eq:zeromomentum}
\gen{S}_{+-}(-\infty,\gamma) \approx
\gen{S}_{+-}(\gamma,+\infty) \approx
\begin{pmatrix}
1   &    \phantom{+}0   &   \phantom{+}0 &  \phantom{+}0 \\
0   &    \phantom{+}0   &   \phantom{+}1 &  \phantom{+}0 \\
0   &    \phantom{+}1   &   \phantom{+}0 &  \phantom{+}0 \\
0   &    \phantom{+}0   &   \phantom{+}0 &  -1 
\end{pmatrix} = \Pi^g
\,,
\end{equation}
which is the graded permutation. If the S~matrix is appropriately normalised, we would expect the right-hand-side to be precisely~$\Pi^g$. In our conventions for the S~matrix, this is precisely how we would describe a trivial scattering, \textit{i.e.}\ a mere reordering of particles. Since the near-BMN expansion is a small-momentum expansion, eq.~\eqref{eq:zeromomentum} also provides the leading (trivial) order of the perturbative expansion. Consistently this gives a free theory.
In a similar way we may check that
\begin{equation}
    \gen{S}_{+\bullet}(-\infty,p) \approx \gen{S}_{\bullet-}(p,+\infty) \approx \Pi^g\,.
\end{equation}
All these formulae hold up to a dressing factor. It will be therefore important to choose such a factor  that the formulae become valid for the ``dressed'' S~matrix too.
Things are a little more involved when the zero-momentum particle appears in other blocks, which will generally be the case, in particular in the Bethe-Yang equations. We find that in this case the matrix part of the S~matrix differs from $\Pi^g$ by a Fermion sign matrix. Such additional sign is necessary to satisfy (among other things) the Yang-Baxter equation. This is most simply expressed when we consider the full S~matrix --- the one arising for the representations of the algebra~\eqref{eq:bigalgebra}. With a slight abuse  of notation, we indicate that $16\times 16$ matrix by $\gen{S}_{\pm\pm}$ too. We have
\begin{equation}
    \gen{S}_{++}(-\infty,\gamma_2)=\Pi^g\,(-1)^{\epsilon_1}\,,\qquad
    \gen{S}_{--}(\gamma_1,+\infty)=\Pi^g\,(-1)^{\epsilon_2}\,,
\end{equation}
where $\epsilon_j$ is the zero if the $j$-th particle is a Boson, and it is one if it is a Fermion.%
\footnote{
In terms of the tensor-product decomposition of the two-particle state, $(-1)^{\epsilon_1}$ is given by the matrix $\Sigma\otimes\gen{1}$, while $(-1)^{\epsilon_1}$ is $\gen{1}\otimes\Sigma$
.}
It would be interesting to investigate the effect of this additional sign on the Bethe-Yang equations.

\paragraph{Coincident rapidities.}
Another special configuration, which plays a role in particular in the derivation of the algebraic Bethe ansatz, is that of identical particles that have coincident rapidities. We can easily verify that
\begin{equation}
    \gen{S}_{\pm\pm}(\gamma,\gamma) \approx
\begin{pmatrix}
1   &    0   &   0 &  0 \\
0   &    1   &   0 &  0 \\
0   &    0   &   1 &  0 \\
0   &    0   &   0 &  1 
\end{pmatrix} = \gen{1}\,.
\end{equation}
If the S~matrix is appropriately normalised, we would expect the right-hand side to be a constant $c\gen{1}$.
In the Zamolodchikov-Faddeev algebra, any $c\neq1$ would yield a constraint preventing solutions with repeated rapidities. We will assume that this is the case.
Notice that an equation of this type does not hold for $\gen{S}_{\pm\mp}(\gamma,\gamma)$, where coincident rapidities do not mean identical particles. Taking the momenta to the same value and then taking them to zero, versus taking them to zero one after the other give different S~matrices. This is expected in general; a typical example of this is the S~matrix of the Heisenberg model, which famously takes the form $S(u_{12})=(u_{12}+2i)/(u_{12}-2i)$ with $u_{12}=\cot \tfrac{p_1}{2}-\cot \tfrac{p_2}{2}$.

\subsection{Constraints on the scalar factors}
Having discussed the matrix part of the S~matrix, we now want to discuss symmetries that can help us constrain its dressing factors. It is important to remark that the constraints on the dressing factors should be imposed on the full S~matrix, whose blocks loosley speaking take the form $\gen{S}\otimes\gen{S}$~\cite{Borsato:2014hja}. For the purpose of identifying the features due to the chiral--anti-chiral split it will be sufficient and easier to consider a single block of the S~matrix.

\paragraph{Physical and generalised unitarity.}
A first requirement is that, when the particles scattering are physical and have real momentum, the S~matrix is a unitary matrix. In its most general form this reads
\begin{equation}
    \gen{S}(p_1,p_2)^\dagger\,\gen{S}(p_1,p_2)=\gen{1}\,,\qquad
    v_1>v_2,\qquad p_1,p_2\in\mathbb{R}\,,
\end{equation}
where $\gen{S}(p_1,p_2)$ may be any of the S~matrices that we discussed. By imposing this condition for the various cases , we find constraints on the dressing factors,%
\begin{equation}
    |\Sigma_{\alpha\beta}(\gamma_1,\gamma_2)|=|\Sigma_{\alpha\bullet}(\gamma,p)|= |\Sigma_{\bullet\alpha}(p,\gamma)|=1\,,\qquad
    v_1>v_2,\qquad p_1,p_2\in\mathbb{R}\,.
\end{equation}
A stronger condition is \textit{generalized} physical unitarity, which holds for arbitrary values of the momenta and reads
\begin{equation}
    \gen{S}(p_1^*,p_2^*)^\dagger\,\gen{S}(p_1,p_2)=\gen{1}\,.
\end{equation}
While there can be some subtlety in imposing this condition in massless relativistic theories (see for instance~\cite{Zamolodchikov:1992zr} and appendix~A in~\cite{Baggio:2017kza}), we can see from direct inspection that the matrix part of the S~matrices that we described satisfies this stronger condition. It appears natural therefore to require
\begin{equation}
    \Sigma_{\alpha\beta}(\gamma_{1}^*,\gamma_2^*)^*\,\Sigma_{\alpha\beta}(\gamma_{1},\gamma_2)=
    \Sigma_{\alpha\bullet}(\gamma^*,p^*)^*\,\Sigma_{\alpha\bullet}(\gamma,p)=
    \Sigma_{\bullet\alpha}(p^*,\gamma^*)^*\,\Sigma_{\bullet\alpha}(p,\gamma)=1\,.
\end{equation}

\paragraph{Braiding unitarity.}
Braiding unitarity emerges as a consistency condition of the Zamolod-chikov-Faddeev algebra, see for instance~\cite{Arutyunov:2009ga} for a review, and reads
\begin{equation}
    \gen{S}(p_1,p_2)\,\gen{S}(p_2,p_1)=\gen{1}\,.
\end{equation}
By specialising this relation to the different S-matrices, we find that it reduces to a condition on the dressing factors
\begin{equation}
\label{eq:braiding}
    \Sigma_{\alpha\beta}(\gamma_1,\gamma_2)\,\Sigma_{\beta\alpha}(\gamma_2,\gamma_1)=
    \Sigma_{\alpha\bullet}(\gamma,p)\,\Sigma_{\bullet\alpha}(p,\gamma)=1\,.
\end{equation}
While this condition is a constraint on $\Sigma_{++}(\gamma)$, as well as on $\Sigma_{--}(\gamma)$, it relates the remaining dressing factors to one another. For instance, we can think of it as a way to define $\Sigma_{-+}(\gamma)$ in terms of $\Sigma_{+-}(\gamma)$, and so on.

\paragraph{Parity.}
Another transformation that we can use to relate different blocks of the S~matrix is the parity transformation, which clearly will relate anti-chiral and chiral particles. In particular we have
\begin{equation}
    \Pi^g\cdot\gen{S}(-p_1,-p_2)\cdot\Pi^g = \big(\gen{S}(p_1,p_2)\big)^{-1}\,,
\end{equation}
where $\Pi^g$ is the graded permutation matrix introduced above.
Written down for the massless-massless scattering, this relation takes the form
\begin{equation}
\begin{aligned}
    \Pi^g\cdot\gen{S}_{--}(-\gamma_1,-\gamma_2)\cdot\Pi^g= \big(\gen{S}_{++}(\gamma_1,\gamma_2)\big)^{-1}\,,\\
    \Pi^g\cdot\gen{S}_{-+}(-\gamma_1,-\gamma_2)\cdot\Pi^g= \big(\gen{S}_{+-}(\gamma_1,\gamma_2)\big)^{-1}\,.
\end{aligned}
\end{equation}
For the scattering of same-chirality particles we hence get a new constraint,
\begin{equation}
    \Sigma_{++}(\gamma_1,\gamma_2)\,\Sigma_{--}(-\gamma_1,-\gamma_2)=1\,,
\end{equation}
which taken together with the braiding unitarity condition~\eqref{eq:braiding} gives
\begin{equation}
\label{eq:samechiralityidentity}
    \Sigma_{++}(\gamma_1,\gamma_2) = \Sigma_{--}(-\gamma_2,-\gamma_1)\,.
\end{equation}
For particles of opposite chirality we get instead
\begin{equation}
    \Sigma_{+-}(\gamma_1,\gamma_2)\,\Sigma_{-+}(-\gamma_1,-\gamma_2)=1\,,
\end{equation}
which together with braiding unitarity~\eqref{eq:braiding} gives
\begin{equation}
    \Sigma_{\pm\mp}(\gamma_1,\gamma_2) = \Sigma_{\pm\mp}(-\gamma_2,-\gamma_1)\,.
\end{equation}
For mixed-mass scattering instead we get
\begin{equation}
    \Pi^g\cdot\gen{S}_{\bullet-}(-p,-\gamma)\cdot\Pi^g = \big(\gen{S}_{\bullet+}(p,\gamma)\big)^{-1},\qquad
    \Pi^g\cdot\gen{S}_{-\bullet}(-\gamma,-p)\cdot\Pi^g = \big(\gen{S}_{+\bullet}(\gamma,p)\big)^{-1},
\end{equation}
which reduces to the following equations for the dressing factors
\begin{equation}
    \Sigma_{\bullet-}(-p,-\gamma)\,\Sigma_{\bullet+}(p,\gamma)=1\,,\qquad
    \Sigma_{-\bullet}(-\gamma,-p)\,\Sigma_{+\bullet}(\gamma,p)=1\,.
\end{equation}
Combining these conditions with those stemming from braiding unitarity, we find that the mixed-mass dressing factors can be entirely determined in terms of a single block, for instance~$\Sigma_{+\bullet}(\gamma,p)$.

\paragraph{Time reversal.}
Due to time-reversal invariance we expect, for real momenta
\begin{equation}
    \gen{S}(-p_1,-p_2)^* = \gen{S}(p_1,p_2)\,.
\end{equation}
Spelling out the equation we get
\begin{equation}
    \gen{S}_{++}(\gamma_1,\gamma_2)^*=\gen{S}_{--}(-\gamma_1,-\gamma_2)\,,\qquad
    \gen{S}_{+-}(\gamma_1,\gamma_2)^*=\gen{S}_{-+}(-\gamma_1,-\gamma_2)\,,
\end{equation}
and
\begin{equation}
    \gen{S}_{+\bullet}(\gamma,p)^*=\gen{S}_{-\bullet}(-\gamma,-p)\,,\qquad
    \gen{S}_{\bullet+}(p,\gamma)^*=\gen{S}_{\bullet-}(-p,-\gamma)\,,
\end{equation}
which is compatible with all the previous constraints and with the matrix structure. This does not result in new constraints on the dressing factors.

\paragraph{Crossing symmetry.}
Crossing symmetry  was originally understood as a property of relativistic quantum field theories. However, it can also be understood in the more-general framework of non-relativistic theory by considering a transformation that flips the sign of energy, momentum and of the abelian central charge $\gen{M}$:
\begin{equation}
    E_c= -E\,,\qquad
    p_c= -p\,,\qquad
    M_c = -M\,,
\end{equation}
while also reversing the notion of highest/lowest weight states.%
\footnote{%
This can be seen by realising crossing symmetry as scattering with a charge-singlet composed by two excitations like in the appendix~D of~\cite{Beisert:2005tm}; necessarily, in the singlet highest- and lowest-weight states will be paired up. See also~\cite{Sfondrini:2014via} for a detailed discussion of this construction in AdS3/CFT2.
}
Not only such a representation exists, but it is also possible to construct it by analytically continuing the representations we considered thus far and acting with a suitably-defined charge conjugation matrix, see cite~\cite{Borsato:2014hja}. Without discussing the details, we denote the analytic continuation $p\to\bar{p}$. It is important to stress that, while $p$ does change sign under crossing, we are not dealing with a parity transformation. In fact, here we also want $E$ (and $M$) to change sign. If we describe $(p,E)$ in terms of $\gamma$ this can be achieved by redefining the argument of both functions as
\begin{equation}
    \gamma\to \gamma+i\pi\,,
\end{equation}
and defining
\begin{equation}
    E_c=E(\gamma+i\pi)=-E(\gamma)\,,\qquad
    p_c=p(\gamma+i\pi)=-p(\gamma)\,,
\end{equation}
which behaves as we want for real $\gamma_c$. This does not change the chirality of the particles, as that depends on $\partial E/\partial p$. Using the Zamolodchikov-Faddeev algebra we can derive~\cite{Arutyunov:2009ga} an invariance condition of the S~matrix which reads, in general
\begin{equation}
\label{eq:crossingeq}
    \gen{c}\otimes\gen{1}\cdot \left(\Pi^g\,\gen{S}(p_{1,c},p_2)^{t_1}\right)\cdot\gen{c}^{-1}\otimes\gen{1}\cdot\Pi^g\gen{S}(p_1,p_2)=\gen{1}\,,
\end{equation}
where $\gen{c}$ is the charge-conjugation matrix and $t_1$ denotes transposition in the first space.
This equation has been considered in~\cite{Borsato:2014hja} for the full S~matrix $\gen{S}\otimes\gen{S}$. It holds separately for all the various dressing factors we considered, given that crossing does not mix different chiralities.
We conclude our discussion of crossing with one important caveat: the charge-conjugation matrix $\gen{c}$ should, in particular, exchange the highest-weight and lowest-weight state. If we tried to define such an operation of the two-dimensional representation of~\eqref{eq:representations} we would have that $\gen{c}$ exchanges Bosons with Fermions (or, in other words, that $\gen{c}$ does not commute with $\Sigma$). This is highly unusual and in fact incompatible with the derivation of~\eqref{eq:crossingeq}. In fact, we should always demand
\begin{equation}
\label{eq:chargeconj}
    \Sigma\,\gen{c}\,\Sigma=\gen{c}\,.
\end{equation}
It is impossible to construct such a charge-conjugation matrix for the two-dimensional representations~\eqref{eq:representations}. However, no issue arises when we consider the full S~matrix which features the tensor product of two representations of the type~\eqref{eq:representations}. Since crossing symmetry is only a requirement on the full S~matrix, we encounter no contradiction.%
\footnote{%
We have recalled earlier that the two-dimensional  representations~\eqref{eq:representations} also appear in $AdS_3\times S^3\times S^3\times S^1$, and that there they (rather than their twofold tensor products) are the fundamental object of the theory. In that case, we see that the crossing transformation relates pair of distinct irreducible representations, one with Bosonic highest weight and one with Fermionic highest weight, so that eq.~\eqref{eq:chargeconj} holds, see~\cite{Borsato:2012ud,Borsato:2015mma}.
}
We refer the reader to~\cite{Sfondrini:2014via, Borsato:2014hja} for a discussion of crossing for $AdS_3\times S^3\times T^4$.

\paragraph{Special values of the momentum.}
We have seen that the S~matrix assumes a particularly simple form when  momenta take some special values. One case was that of coincident momenta. In that case we argued that whenever the S~matrix is not precisely the identity, the Zamolodchikov-Faddeev algebra would rule out the existence of states with coincident momenta. The other interesting case occurs when one of the particles has zero momentum. It should be stressed that, in order to set the correct normalisation for the S~matrix, in this case it is particularly important to work with the full S~matrix (\textit{i.e.}, with $\gen{S}\otimes\gen{S}$). For the case of the physical S~matrix   $\gen{S}_{+-}$ we can demand
\begin{equation}
    \Sigma_{+-}(-\infty,\gamma_2)=\Sigma_{+-}(\gamma_1,+\infty)=1\,.
\end{equation}
For the purpose of writing the Bethe-Yang equations we will have to consider the case where one of the momenta is zero in $\gen{S}_{\pm\pm}$. There we have
\begin{equation}
    \Sigma_{++}(-\infty,\gamma_2)=\Sigma_{--}(\gamma_1,+\infty)=1\,.
\end{equation}
Reassuringly, this is compatible with eq.~\eqref{eq:samechiralityidentity}.

\section{Mirror model}
\label{sec:mirror}
So far we have discussed the theory emerging from the string non-linear sigma model. We now want to turn to the mirror theory, again with special attention to the case of massless particles.

\subsection{Kinematics of the mirror theory}
The mirror theory is related to the original one by a double Wick rotation~\cite{Zamolodchikov:1989cf}. This is rather harmless for relativistic theories but it results in a completely new model in a non-relativistic setup~\cite{Arutyunov:2007tc}. In practice, we define
\begin{equation}
    E_m = -i\,p\,,\qquad p_m = -i\, E\,,
\end{equation}
so that the dispersion relation for a particle of mass $|M|$, which we recall is
\begin{equation}
    E^2 = M^2+4h^2\sin^2\left(\frac{p}{2}\right)\,,
\end{equation}
becomes
\begin{equation}
    -p_m^2 = M^2-4h^2\sinh^2\left(\frac{E_m}{2}\right)\,,
\end{equation}
or in other words
\begin{equation}
\label{eq:mirrordispersion}
    E_m(p_m,M) = 2\arcsinh\left(\frac{\sqrt{M^2+p_m^2}}{2h}\right)\,.
\end{equation}
When $|M|\geq1$ this is the same kinematics as for $AdS_5\times S^5$, which was discussed at length in~\cite{Arutyunov:2007tc}. Once again the new and more interesting case is that of $M=0$ which gives
\begin{equation}
    E_m(p_m,0) = 2\arcsinh\left(\frac{|p_m|}{2h}\right)\,,
\end{equation}
which is once again apparently non-analytic.
It is worth noting that the group velocity
\begin{equation}
    v_m=\frac{2p_m}{\sqrt{M^2+p_m^2}\sqrt{M^2+p^2_m+4h^2}}\,,
\end{equation}
is still non-relativistic also when $M=0$. Like before, the zero-momentum limit is special. The corresponding massless particles have zero-energy and their velocity is maximal in module $v_m=\pm1/h$.%
\footnote{%
Unlike what happens in string theory, massive particle at zero (mirror) momentum do not have constant, \textit{i.e.}\ coupling independent, energy. This is because the mirror model does not have global (super)isometries whose descendants may be created by adding zero-mirror-momentum particles. This can be seen by looking at the geometry of the mirror model~\cite{Arutyunov:2014cra, Arutyunov:2014jfa}.
}

\subsection{Symmetry algebra and representations in the mirror theory}
Starting from symmetry algebra of the original theory, it is convenient to consider the following linear transformation of the supercharges, cf. \cite{Arutyunov:2007tc}:
 \bal
 \gen{q}_m &= \frac{1}{\sqrt{2}}\big(  \gen{q} +i\, \widetilde{ \gen{s} }  \big)\,,\qquad \gen{s}_m = \frac{1}{\sqrt{2}}\big(  \gen{s} + i\, \widetilde{ \gen{q} }  \big),
 \\
 \widetilde{\gen{q}}_m &= \frac{1}{\sqrt{2}}\big(  \widetilde{ \gen{q} } +i \,  \gen{s}  \big)\,,\qquad \widetilde{\gen{s}}_m = \frac{1}{\sqrt{2}}\big(  \,\widetilde{ \gen{s} } +i \,  \gen{q}  \big).
 \eal
 Notice that this transformation does not preserve the real form of the algebra; this is fine, because the double Wick rotation also does not preserve reality.
 In terms of these new charges, the non-trivial anti-commutation relations are
 \bal
 \label{eq:mirrorrotation}
    \{\gen{q}_m,\gen{s}_m\}&=\frac{1}{2}(-ih\sin\gen{p}+ \gen{M}) = \frac{1}{2}(h\sinh\gen{E}_m+ \gen{M}) = \gen{H}_m\,,\\
    \{\tilde{\gen{q}}_m,\tilde{\gen{s}}_m\}&={\frac{1}{2}}(-ih\sin\gen{p}- \gen{M}) = \frac{1}{2}(h\sinh\gen{E}_m- \gen{M}) =\widetilde{\gen{H}}_m\,,
    \\
    \{\gen{q}_m,\tilde{\gen{q}}_m\}&=\frac{1}{2}(i\gen{E}+ih\cos\gen{p}-ih) = \frac{1}{2}(-\gen{p}_m+ih \cosh\gen{E}_m - i h) = \gen{C}_m\,,
    \\
    \{\gen{s}_m,\tilde{\gen{s}}_m\}&=\frac{1}{2}(i\gen{E}-ih\cos\gen{p}+ih) = \frac{1}{2}(-\gen{p}_m-ih \cosh\gen{E}_m+ih) =\bar{\gen{C}}_m\,,\
 \eal
Note that, for a physical particle of real mirror momentum~$p_m$ and dispersion~\eqref{eq:mirrordispersion}, we find that 
\begin{equation}
    H_m\geq 0\,,\qquad \widetilde{H}_m\geq0\,,\qquad C^*_m=\bar{C}_m\,,
\end{equation}
meaning that this defines a unitary representation.

\paragraph{Shortening condition.}
Once again we can derive a shortening condition for the algebra
\begin{equation}
    \gen{H}_m \widetilde{\gen{H}}_m=\gen{C}_m\bar{\gen{C}}_m\,,
\end{equation}
which in terms of the mirror momentum gives
\begin{equation}
    E_m = \arccosh\left(1+\frac{M^2+p^2_m}{2h^2}\right) = 2\arcsinh\left(\frac{\sqrt{M^2+p^2_m}}{2h}\right)\,,
\end{equation}
confirming that for a massless particle this is not analytic.%
\footnote{%
In fact, $\arccosh(1+z^2)$ is not analytic across the imaginary axis.} 

\paragraph{Central charges.}
We can compute the value of the central charges in terms of the rapidity variable $\gamma$. In fact, it is convenient to redefine
\begin{equation}
\label{eq:mirrorshift}
    \gamma = \gamma_m+\frac{i\pi}{2}\,,
\end{equation}
so that we find that
\begin{equation}
\label{eq:mirrorcentralcharges}
    \sinh E_m=\frac{2\cosh\gamma_m}{\sinh^2\gamma_m}\,,\qquad
    p_m=-\frac{2}{\sinh\gamma_m}\,,
\end{equation}
so that the mirror energy is positive-definite for real values of~$\gamma_m$ and the mirror momentum is real. Trying to derive $E_m(\gamma_m)$ yields a non-analytic function once again: formally we get
\begin{equation}
    E_m(\gamma_m)=\log\frac{(1+e^{\gamma_m})^2}{(1-e^{\gamma_m})^2}\,,\qquad \gamma_m\in\mathbb{R}\,,
\end{equation}
which is non-analytic across the imaginary-$\gamma_m$ axis.
To avoid this problem we can start from the definition of the momentum for anti-chiral and chiral particles which we gave in the string-theory region. Using that, we define for anti-chiral particles
\begin{equation}
    \text{anti-chiral}:\quad E_m(\gamma_m) = -2\log\frac{1-e^{\gamma_m}}{1+e^{\gamma_m}}+2\pi i\,,\qquad \gamma_m>0\,,\quad p_m<0\,,
\end{equation}
while for chiral particles
\begin{equation}
    \text{chiral}:\quad E_m(\gamma_m) = -2\log\frac{1-e^{\gamma_m}}{1+e^{\gamma_m}}\,,\qquad\qquad \gamma_m<0\,,\quad p_m>0\,.
\end{equation}

\paragraph{Representations.}
We define the new representation coefficients
\begin{equation}
\begin{aligned}
    \gen{q}_{m,\pm}=\begin{pmatrix}
    0&0\\
    a_{m,\pm}&0
    \end{pmatrix},
    \qquad
    \gen{s}_{m,\pm}=\begin{pmatrix}
    0&a_{m,\pm}^*\\
    0&0
    \end{pmatrix},\\
    \tilde{\gen{s}}_{m,\pm}=\begin{pmatrix}
    0&0\\
    b_{m,\pm}^*&0
    \end{pmatrix},\qquad
    \tilde{\gen{q}}_{m,\pm}=\begin{pmatrix}
    0&b_{m,\pm}\\
    0&0
    \end{pmatrix},
\end{aligned}
\end{equation}
where the form of the representation coefficients will follow from the rotation~\eqref{eq:mirrorrotation} and from the redefinition~\eqref{eq:mirrorshift}.
For anti-chiral representations we find
\begin{equation}
    a_{m,-}=b_{m,-}=\sqrt{2h}e^{\gamma_m/2}\frac{e^{\gamma_m}+i}{e^{2\gamma_m}-1}\,,\qquad
    a_{m,-}^*=b_{m,-}^*=\sqrt{2h}e^{\gamma_m/2}\frac{e^{\gamma_m}-i}{e^{2\gamma_m}-1}\,,
\end{equation}
while for the chiral representation we find
\begin{equation}
    a_{m,+}=-b_{m,+}=\frac{\sqrt{2h}}{-i}e^{\gamma_m/2}\frac{e^{\gamma_m}+i}{e^{2\gamma_m}-1}\,,\qquad
    a_{m,+}^*=-b_{m,+}^*=\frac{\sqrt{2h}}{i}e^{\gamma_m/2}\frac{e^{\gamma_m}-i}{e^{2\gamma_m}-1}\,.
\end{equation}

\paragraph{On the two-particle representation.}
By itself, the fact that we encounter different expressions for the one-particle representation is not an issue, as we can use a change of the one-particle basis to relate the chiral case to the antichiral. Like before the important question is what happens in the two-particle representation. Here we encounter a coproduct featuring
\begin{equation}
    e^{- E_m/2}=\exp\left(-\frac{1}{2}\log\frac{(1+e^{\gamma_m)^2}}{(1-e^{\gamma_m)^2}}\right),
\end{equation}
where we used the non-analytic expression derived from~\eqref{eq:mirrorcentralcharges} for the mirror energy. Despite the exponential function, the same factor of one-half that caused troubles earlier --- making the two-particle string representation non-periodic --- makes this expression non-analytic. We can resolve this issue by splitting the representation in  anti-chiral and chiral parts, whereby we find
\begin{equation}
    e^{- E_m/2} =
    \begin{cases}
    +\tanh\displaystyle\frac{\gamma_m}{2}\,,\qquad&\gamma_m>0\qquad\text{(anti-chiral)}\,,\\[0.3cm]
    -\tanh\displaystyle\frac{\gamma_m}{2}\,,\qquad&\gamma_m<0\qquad\text{(chiral)}\,.
    \end{cases}
\end{equation}
Therefore, a degree of caution is needed when describing the different branches of the mirror theory, like for the string worldsheet theory.
Sector by sector, the mirror S~matrix will take the same form of the string S~matrix, with the only difference that its rapidity variables will have to be continued to the mirror region.

\section*{Acknowledgements}
AS thanks Nima Arkani-Hamed, Hofie Hannesdottir and Sebastian Mizera for interesting related discussions. AS gratefully acknowledges support from the IBM Einstein Fellowship.

\appendix

\section{Explicit form for the mixed-mass S~matrix}
\label{app:smatrix}
The mixed-mass S~matrix is not of difference form, and the kinematics of the massive particle depend on $x^\pm$. It follows the construction of ref.~\cite{Borsato:2014hja} and can be found by using the formulae reported there, having care of specifying that, when $M=0$, $x_p=\pm e^{ip/2}$ where the sign depends on whether we are dealing with chiral or anti-chiral massless particles. This can be done when the massive particle has $M=+1$ or $M=-1$. Without detailing all the cases, here we report how the signs due to chiral/anti-chiral particles appear in the case where the massive particle has $M=+1$. We write the S~matrix in the basis $(\ket{\phi},\ket{\varphi})$ for both representations, and assuming for definiteness that in both cases the highest-weight state is a Boson. We find:
\begin{equation}
\begin{aligned}
\gen{S}_{+\bullet}(\gamma,p) &=
\Sigma_{+\bullet}(\gamma,p)\\
&\times
\begin{pmatrix}
1   &    0   &   0 &  0 \\
0   &    \frac{2\sqrt{2}e^{\gamma/2}\eta_p}{\sqrt{h}(e^\gamma+i+(e^\gamma -i)x^+_p)}   &   e^{\tfrac{i}{2}p}\frac{e^\gamma+i+(e^\gamma -i)x^-_p}{e^\gamma+i+(e^\gamma -i)x^+_p} &  0 \\
0   &   -\frac{e^\gamma-i+(e^\gamma +i)x^+_p}{e^\gamma+i+(e^\gamma -i)x^+_p}  &   \frac{2\sqrt{2}e^{\gamma/2}\eta_p}{\sqrt{h}(e^\gamma+i+(e^\gamma -i)x^+_p)} &  0 \\
0   &    0   &   0 &  e^{\tfrac{i}{2}p}\frac{e^\gamma-i+(e^\gamma +i)x^-_p}{e^\gamma+i+(e^\gamma -i)x^+_p}
\end{pmatrix}
\end{aligned}
\end{equation}
\begin{equation}
\begin{aligned}
\gen{S}_{-\bullet}(\gamma,p) &=
\Sigma_{-\bullet}(\gamma,p)\\
&\times
\begin{pmatrix}
1   &    0   &   0 &  0 \\
0   &    \frac{2\sqrt{2}i\,e^{\gamma/2}\eta_p}{\sqrt{h}(e^\gamma+i+(e^\gamma -i)x^+_p)}   &   e^{\tfrac{i}{2}p}\frac{e^\gamma+i+(e^\gamma -i)x^-_p}{e^\gamma+i+(e^\gamma -i)x^+_p} &  0 \\
0   &   \frac{e^\gamma-i+(e^\gamma +i)x^+_p}{e^\gamma+i+(e^\gamma -i)x^+_p}  &   \frac{2\sqrt{2}i\,e^{\gamma/2}\eta_p}{\sqrt{h}(e^\gamma+i+(e^\gamma -i)x^+_p)} &  0 \\
0   &    0   &   0 &  -e^{\tfrac{i}{2}p}\frac{e^\gamma-i+(e^\gamma +i)x^-_p}{e^\gamma+i+(e^\gamma -i)x^+_p}
\end{pmatrix}
\end{aligned}
\end{equation}
\begin{equation}
\begin{aligned}
\gen{S}_{\bullet+}(p,\gamma) &=
\Sigma_{\bullet+}(p,\gamma)\\
&
\begin{pmatrix}
1   &    0   &   0 &  0 \\
0   &    \frac{2\sqrt{2}e^{-ip/2}e^{\gamma/2}\eta_p}{\sqrt{h}(e^\gamma-i+(e^\gamma +i)x^-_p)}   &   -\frac{e^\gamma+i+(e^\gamma -i)x^-_p}{e^\gamma-i+(e^\gamma +i)x^-_p} &  0 \\
0   &   e^{-\tfrac{i}{2}p}\frac{e^\gamma-i+(e^\gamma +i)x^+_p}{e^\gamma-i+(e^\gamma +i)x^-_p}  &   \frac{2\sqrt{2}e^{-ip/2}e^{\gamma/2}\eta_p}{\sqrt{h}(e^\gamma-i+(e^\gamma +i)x^-_p)} &  0 \\
0   &    0   &   0 &  e^{-\tfrac{i}{2}p}\frac{e^\gamma+i+(e^\gamma -i)x^+_p}{e^\gamma-i+(e^\gamma +i)x^-_p}
\end{pmatrix}
\end{aligned}
\end{equation}
\begin{equation}
\begin{aligned}
\gen{S}_{\bullet-}(p,\gamma) &=
\Sigma_{\bullet-}(p,\gamma)\\
&\times
\begin{pmatrix}
1   &    0   &   0 &  0 \\
0   &    \frac{-2\sqrt{2}ie^{-ip/2}e^{\gamma/2}\eta_p}{\sqrt{h}(e^\gamma-i+(e^\gamma +i)x^-_p)}   &   \frac{e^\gamma+i+(e^\gamma -i)x^-_p}{e^\gamma-i+(e^\gamma +i)x^-_p} &  0 \\
0   &   e^{-\tfrac{i}{2}p}\frac{e^\gamma-i+(e^\gamma +i)x^+_p}{e^\gamma-i+(e^\gamma +i)x^-_p}  &   \frac{-2\sqrt{2}ie^{-ip/2}e^{\gamma/2}\eta_p}{\sqrt{h}(e^\gamma-i+(e^\gamma +i)x^-_p)} &  0 \\
0   &    0   &   0 &  -e^{-\tfrac{i}{2}p}\frac{e^\gamma+i+(e^\gamma -i)x^+_p}{e^\gamma-i+(e^\gamma +i)x^-_p}
\end{pmatrix}
\end{aligned}
\end{equation}

\bibliographystyle{JHEP}
\bibliography{refs}
\end{document}